\documentclass[aps,pra,twocolumn,10pt,superscriptaddress,longbibliography]{revtex4-1}

\usepackage[utf8]{inputenc}
\usepackage[T1]{fontenc}
\usepackage[sc,osf]{mathpazo}
\usepackage{amsmath, amsthm, amssymb}
\usepackage{graphicx}
\usepackage{dcolumn}
\usepackage{bm}
\usepackage{bbm}
\usepackage{color}
\usepackage[usenames,dvipsnames]{xcolor}
\usepackage{hyperref}
\usepackage{nicefrac}

\newcommand{\eqnref}[1]{Eq.~(\ref{#1})}
\newcommand{\eref}[1]{(\ref{#1})}
\newcommand{\figref}[1]{Fig.~\ref{#1}}

\newcommand{\secref}[1]{Sec.~\ref{#1}}
\newcommand{\appref}[1]{Appendix~\ref{#1}}

\newcommand{\citeref}[1]{Ref.~\cite{#1}}

\newcommand{\sh}{\hat \sigma}
\newcommand{\ket}[1]{| #1 \rangle}

\renewcommand{\t}[1]{\textrm{#1}}

\newcommand{\ii}{\mathrm{i}}
\newcommand{\ee}{\mathrm{e}}

\DeclareGraphicsExtensions{.png,.pdf}

\begin{document}
\title{Improved Quantum Magnetometry beyond the Standard Quantum Limit}

\author{J. B. Brask}\affiliation{D\'epartement de Physique Th\'eorique, Universit\'e de Gen\`eve, 1211 Geneva, Switzerland}
\author{R. Chaves}\affiliation{Institute for Physics, University of Freiburg, Rheinstrasse 10, D-79104 Freiburg, Germany}\affiliation{Institute  for  Theoretical  Physics,  University  of  Cologne, 50937 Cologne,  Germany}
\author{J. Ko\l{}ody\'{n}ski}\affiliation{ICFO--Institut de Ci\`encies Fot\`oniques, Mediterranean Technology Park, 08860 Castelldefels (Barcelona), Spain}\affiliation{Faculty of Physics, University of Warsaw, 02-093 Warsaw, Poland}

\begin{abstract}
Under ideal conditions, quantum metrology promises a precision gain over classical techniques scaling quadratically with the number of probe particles. At the same time, no-go results have shown that generic, uncorrelated noise limits the quantum advantage to a constant factor. In frequency estimation scenarios, however, there are exceptions to this rule and, in particular, it has been found that transversal dephasing does allow for a scaling quantum advantage. Yet, it has remained unclear whether such exemptions can be exploited in practical scenarios. Here, we argue that the transversal-noise model applies to the setting of recent magnetometry experiments and show that a scaling advantage can be maintained with one-axis-twisted spin-squeezed states and Ramsey-interferometry-like measurements. This is achieved by exploiting the geometry of the setup that, as we demonstrate, has a strong influence on the achievable quantum enhancement for experimentally feasible parameter settings. When, in addition to the dominant transversal noise, other sources of decoherence are present, the quantum advantage is asymptotically bounded by a constant, but this constant may be significantly improved by exploring the geometry.
\end{abstract}

\maketitle

\section{Introduction}

High-precision parameter estimation is fundamental throughout science. Quite generally, a number of probe particles are prepared, then subjected to an evolution which depends on the quantity of interest, and finally measured. From the measurement results an estimate is then extracted. When the particles are classically correlated and non-interacting, as a consequence of the central limit theorem, the mean-squared error of the estimate decreases as $1/N$, where $N$ is the number of particles (probe size). This best scaling achievable with a classical probe is known as the \emph{standard quantum limit} (SQL) \cite{Giovannetti2004}. Quantum metrology aims to improve estimation by exploiting quantum correlations in the probe.

In an ideal setting without noise, it is well known that quantum resources allow for a quadratic improvement in precision over the SQL~\cite{Giovannetti2004,Giovannetti2006}; i.e, 
the mean-squared error of the estimate after a sufficient number of experimental repetitions can scale as $1/N^2$, yielding the the so-called \emph{Heisenberg limit}. Realistic evolution, however, always involves noise of some form, and although quantum metrology has been demonstrated experimentally, e.g., for atomic magnetometry \cite{Wasilewski2010,Koschorreck2010,Sewell2012,Ockeloen2013,Sheng2013,Lucivero2014,Muessel2014}, spectroscopy \cite{Meyer2001,Leibfried2004}, and clocks \cite{Appel2009,Louchet2010}, there is currently much effort to determine exactly when, and by how much, quantum resources allow estimation to be improved in the presence of decoherence \cite{Huelga1997,Ulam2001,Shaji2007,Fujiwara2008,Ji2008,Hayashi2011,Demkowicz2012,Kolodynski2013,Kolodynski2014,Demkowicz2014,Escher2011,Matsuzaki2011,Chin2012,Chaves2013,Kessler2014,Duer2014,Arrad2014,Froewis2014,Knysh2014,Jarzyna2014}. It is known that for most types of uncorrelated noise (acting independently on each probe particle) the asymptotic scaling is constrained to be SQL-like \cite{Fujiwara2008,Ji2008,Hayashi2011,Escher2011,Demkowicz2012,Kolodynski2013,Kolodynski2014,Demkowicz2014}. Specifically, when estimating a parameter $\omega$, the mean-squared error obeys $\Delta^2\omega \geq r/\nu N$, where $\nu$ is the number of repetitions and $r$ is a constant which depends on the evolution. If the evolution, which each probe particle undergoes, is independent of $N$, the scaling is constrained to be SQL-like. However, for frequency estimation this is not necessarily the case. In frequency estimation scenarios, such as those of atomic  magnetometry \cite{Auzinsh2004,Kominis2008,Liu2015}, spectroscopy \cite{Wineland1992,Wineland1994,Bollinger1996}, and clocks \cite{Buzek1999,Andre2004,Macieszczak2014,Ostermann2013,Borregaard2013,Kessler2014a}, there are two relevant resources, the total number of probe particles $N$ and the total time $T$ available for the experiment \cite{Huelga1997,Ulam2001}. The experimenter is free to choose the interrogation time $t \!=\! T/\nu$ and, in particular, $t$ may be adapted to $N$. In this case, the time over which unitary evolution and decoherence act is different for each $N$ and thus the evolution is not independent of $N$. Schematically, the no-go results for noisy evolution in this case become
\begin{equation}
\label{eq.SQLlike}
\Delta^2 \omega \geq \frac{r(t)}{N T/t} \quad \underset{t(N)}{\xrightarrow{\hspace*{1cm}}} \quad \Delta^2\omega T \geq \frac{c(t(N))}{N} ,
\end{equation}
with $c(t) \!=\! r(t)\, t$. Thus, if for some optimal choice of $t(N)$ the coefficient $c$ decreases with $N$, although the no-go results may hold for any fixed evolution time, the bound does not imply SQL-like scaling. Note that the bound \eqnref{eq.SQLlike} is always achievable in the many-repetitions limit $\nu\!\to\!\infty$ \cite{Giovannetti2006}, which corresponds to $T\!\gg\!t$ \cite{Huelga1997,Ulam2001,Shaji2007}. Although without noise it is optimal to take $t$ as large as possible, i.e, $t=T$, for any noisy evolution the optimal $t$ becomes finite because of noise dominating at large times. So the many-repetitions regime can always be ensured by considering sufficiently large $T$ \cite{Escher2011,Matsuzaki2011,Chin2012,Chaves2013}.

In frequency estimation scenarios, for the asymptotic scaling to be superclassical, $c$ must vanish as $N\!\to\!\infty$, which is only possible if the evolution is such that decoherence can be neglected at short time scales, and the no-go theorems then do not apply \cite{Kolodynski2014a}. This is also necessary for error-correction techniques, which utilise ancillary particles not sensing the parameter \cite{Duer2014} or employ correcting pulses during the evolution \cite{Kessler2014}, to surpass the SQL \cite{Kolodynski2014,Demkowicz2014,Kolodynski2014a}. Without such additional resources---considering just interrogation-time optimisation---the possibility of superclassical scaling has been demonstrated for non-Markovian \cite{Matsuzaki2011,Chin2012} evolutions (for which the effective decoherence strength vanishes as $t\!\to\!0$), as well as for dephasing directed along a direction perpendicular to the unitary evolution \cite{Chaves2013}. In the latter case, it was shown that an optimal variance scaling of $1/N^{5/3}$ can be obtained by choosing $t\!\propto\!1/N^{1/3}$ \cite{Chaves2013}. This result was based on numerical analysis of the \emph{quantum Fisher information} (QFI) \cite{Braunstein1994} and was shown to be saturable by Greenberger-Horne-Zeilinger (GHZ) states \cite{GHZ}. However, GHZ states of many particles are not easily generated in practice, and the Fisher information approach does not explicitly provide the required measurements. Thus, the question of whether the scaling is achievable in practically implementable metrology was left open.

In this paper, we argue that the transversal-noise model applies to atomic magnetometry, in particular the experimental setting of \cite{Wasilewski2010}, and study the quantum advantage attainable with use of \emph{one axis-twisted spin-squeezed states} (OATSSs) \cite{Ma2011} and \emph{Ramsey-interferometry-like measurements} \cite{Wineland1992,Wineland1994,Bollinger1996}, both of which are accessible with current experimental techniques. We explicitly show that the setup geometry plays an important role for the achievable quantum enhancement. A suboptimal choice leads to a constant factor of quantum enhancement, while superclassical precision scaling can be maintained for a more appropriate choice. We study the enhancement achievable with the numbers of the experiment \cite{Wasilewski2010} and demonstrate the advantage of modifying the geometry. We further consider the case of noise which is not perfectly transversal and find that, although the asymptotic precision scaling is then again SQL-like, the precision may be substantially enhanced by optimising the geometry. As the previous results \cite{Chaves2013} were based on numerics, we also provide an analytical proof of the scaling for GHZ states in \appref{app.ghz_states}.

\section{Model}

We consider a scheme in which $N$ two-level quantum systems are used to sense
a frequency parameter $\omega$ in an experiment of total duration $T$, divided into rounds of interrogation time $t$.
We keep in mind that this can correspond to atomic magnetometry, in which the particles then represent the
atoms with a spin precessing in a magnetic field at a frequency proportional to the field strength.
As in \citeref{Chaves2013}, we describe the noisy evolution by a master equation of Lindblad form
\begin{equation}
\label{eq.master}
\frac{\partial \rho }{\partial t}=\mathcal{H}\!\left( \rho \right) +\mathcal{L}\!\left( \rho \right) .
\end{equation}
Here, $\mathcal{H}\!\left( \rho\right)\!=\! -\ii\!\left[ \hat H,\rho \right]$ is the unitary part of the evolution that encodes the parameter dependence. The Hamiltonian is given by
\begin{equation}
\hat H=\frac{\omega}{2} {\sum_{k=1}^{N}} \hat\sigma_{z}^{(k)}\!,
\end{equation}
where $\hat\sigma_z^{(k)}$ is a Pauli operator acting on the $k$th particle (qubit). The Liouvillian $\mathcal{L}(\rho)$ describes the noise,
which is uncorrelated on different qubits, so that $\mathcal{L}\!=\! {\sum_{k}} \mathcal{L}^{(k)}$, and for a single qubit we have
\begin{equation}
\mathcal{L}^{(k)}\!\rho =\!-\frac{\gamma }{2}\!\left[ \rho - \alpha_{x} \hat\sigma^{(k)}_{x}\!\rho \hat\sigma^{(k)} _{x}\!- \alpha_{y} \hat\sigma^{(k)} _{y}\!\rho \hat\sigma^{(k)} _{y}\!- \alpha_{z} \hat\sigma^{(k)}_{z}\!\rho \hat\sigma^{(k)} _{z} \right]\!,
\label{lindblad}
\end{equation}
where $\gamma$ is the overall noise strength and $\alpha_{x,y,z}\!\geq\! 0$ with $\alpha_{x}\!+\!\alpha_{y}\!+\!\alpha_{z}\!=\!1$. For $\alpha_{z}\!=\!1$,
\eqnref{lindblad} describes dephasing along the direction of the unitary, while $\alpha_{x}\!=\!1$ (or equivalently $\alpha_{y}\!=\!1$) corresponds to the transversal-dephasing noise.
For $\alpha_{x}\!=\!\alpha_{y}\!=\!\alpha_{z}\!=\!1/3$, we have an isotropic depolarizing channel.

Under this model, interrogation-time optimisation leads to a quantum scaling advantage for transversal ($\alpha_x\!=\!1$) but not for parallel ($\alpha_z\!=\!1$) noise.
This can be understood by looking at how the coefficient $c$ in \eqnref{eq.SQLlike} behaves in the two cases. For short times, one can obtain bounds of the form \eqnref{eq.SQLlike} with \cite{Chaves2013}
\begin{eqnarray}
& c_z(\gamma,\omega,t)= 2\gamma +2\gamma^2t + O(t^2), \label{eq.cZ}\\
& c_x(\gamma,\omega,t)= \frac{\gamma^2 \omega^3 t^3}{12}+O(t^5).
\end{eqnarray}
From this, we see that for parallel dephasing, interrogation-time optimisation cannot prevent asymptotic SQL-like scaling, because $c_z$ is bounded from below by the non-zero factor of $2\gamma$. However, for perpendicular noise $c_x \!\rightarrow \!0$ as $t \!\rightarrow \!0$, and hence in this case, optimisation may allow for superclassical scaling. In \citeref{Chaves2013}, it was found that taking $t\!=\!(3 /\gamma \omega^2 N)^{1/3}$ leads to
\begin{equation}
\label{eq.prevbound}
\Delta^2\omega T \;\geq\; \frac{3^{2/3}}{2} (\gamma\omega^2)^{1/3}\, \frac{1}{N^{5/3}},
\end{equation}
and that this bound is achievable with the GHZ input states.

\begin{figure}
\begin{center}
\includegraphics[width=0.99\linewidth]{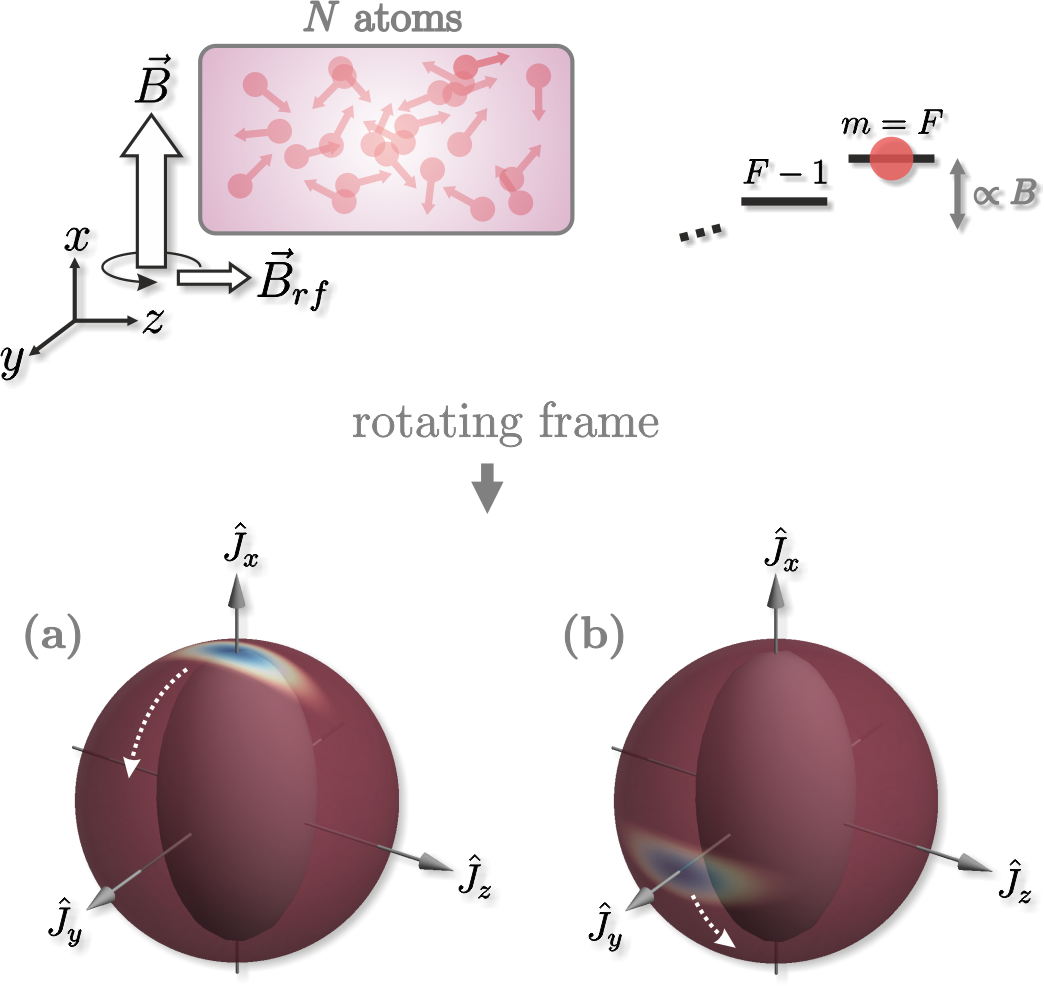}
\end{center}
\caption{\textbf{Atomic magnetometry setup}.
An ensemble of atoms is placed in a strong magnetic field $B$ which induces a level splitting between the magnetic sublevels.
The atoms are used to sense a weak field $B_\t{rf}$ in the plane perpendicular to $B$, which rotates in this plane with a frequency
matched to the Larmor precession induced by $B$. We consider two cases for the state preparation and readout. Scenario (a)
corresponds to the geometry of the experiment \cite{Wasilewski2010}. All atoms are initially pumped to an extreme magnetic sublevel
$m\!=\!F$ creating a coherent spin state aligned with $B$. The state is then squeezed to make it more sensitive to the evolution induced by $B_\t{rf}$.
In a frame rotating around $B$ at the Larmor frequency, the state can be depicted as shown in the lower part. $B_\t{rf}$ points then along $z$ and induces a rotation around the $z$ axis.
The state is squeezed in $y$ and $B_\t{rf}$ is estimated from a measurement of the collective spin component $\hat{J}_y$.
Scenario (b) is similar, but the state is initially perpendicular to $B$. In the rotating frame, it is squeezed in $x$ and $\hat{J}_x$ is measured.
The dominant noise in both cases comes from individual atomic motion causing variations in the effective magnetic field
and hence the energy splitting. This results in uncorrelated dephasing noise in the direction of $B$; the impact on the collective spin
is schematically illustrated by the inner prolate spheroids. Importantly, the noise preserves the spin along $B$ but shrinks it
in the perpendicular directions.}
\label{fig.magnetosetup}
\end{figure}

To see that the model is relevant in practice, we consider the atomic magnetometry experiment of \citeref{Wasilewski2010} illustrated in \figref{fig.magnetosetup}. In this experiment, entanglement was demonstrated to enhance the sensitivity, but the precision scaling with $N$ was not studied. The relevant magnetometer consists of a vapour of caesium atoms, which is subject to a strong external dc magnetic field $B$ and used to sense a weak radio-frequency field $B_\t{rf}$ perpendicular to $B$ (note that in \citeref{Wasilewski2010} two separate ensembles were used; this is not important for the present argument). The atoms are optically pumped into an extreme magnetic sublevel and may be treated as effective two-level systems with an energy splitting determined by $B$. With $B\! \gg\! B_\t{rf}$, the dominant noise is due to small variations in the dc magnetic field seen by different atoms (e.g., due to field inhomogeneities and atomic motion), which leads to fluctuations of the individual energy splittings. This corresponds to a dephasing noise which acts on each atom independently and is characterised by the spin-decoherence time $T_2$ \cite{Wang2005}. As the experiment is conducted at a time scale much shorter than the ones of spontaneous emission and $B$-field fluctuations, other noise sources are suppressed. In particular, the spin-relaxation time $T_1$ can be taken infinite (see \appref{app.master} for discussion) and collective noise can be neglected. The Larmor frequency of the strong field $B$ is matched to the frequency of the weak field $B_\t{rf}$, and it is then convenient to describe the system in a \emph{rotating frame} (RF). If $\rho$ is the state of a single atom in the non-rotating frame and $B$ is directed along the $x$ axis (see \figref{fig.magnetosetup}), the state in the rotating frame reads $\rho_\t{RF} \!=\! \ee^{\ii \hat H_B t} \rho\, \ee^{-\ii \hat H_B t}$, where $\hat H_B \!=\! \kappa B \hat{\sigma}_x$ and $\kappa$ is the coupling strength to the magnetic field. In such a Larmor-precessing frame, the master equation for the evolution may be written as (see also \appref{app.master}) \cite{Kominis2008}
\begin{equation}
\label{eq.master_magnetometry}
\frac{\partial\rho_\t{RF}}{\partial t} = - \ii\kappa B_\t{rf}\left[\hat\sigma_{z},\rho_\t{RF}\right] - \frac{1}{T_{2}}\left(\rho_\t{RF} - \hat\sigma_{x}\rho_\t{RF} \hat\sigma_{x}\right) ,
\end{equation}
where the first term can be understood as the effective free Hamiltonian in the rotating frame with the $B_\t{rf}$ field pointing along $z$. The dephasing noise is directed along $B$ and parametrised by $T_2$. Since \eqnref{eq.master_magnetometry} is exactly of the \eqnref{eq.master} form, it is clear that this experimental setting is captured by the previously stated model with $\omega \!=\! 2\kappa B_\t{rf}$ and transversal noise $\alpha_x \!=\! 1$, $\alpha_y\!=\!\alpha_z\!=\!0$, $\gamma\!=\!2/T_2$. We note that $B \!\gg\! B_\t{rf}$ is important for the noise to be transversal, which may imply that $\gamma$ is large relative to $\omega$. In particular, this is the case in \citeref{Wasilewski2010}, as we show below.

In \citeref{Wasilewski2010}, superclassical precision was demonstrated by initially aligning the collective spin of the atomic ensemble \textit{along} $B$ and reducing fluctuations of its component in the direction perpendicular to both $B$ and $B_\t{rf}$ via spin squeezing [\figref{fig.magnetosetup}(a)]. Below, we study such a geometry along with another setting, in which the collective spin is initially \textit{perpendicular} to both $B$ and $B_\t{rf}$ and its component along $B$ is squeezed [\figref{fig.magnetosetup}(b)]. In principle, scenario (b) can be obtained from (a) by applying a $\pi/2$ pulse to the atomic ensemble before the evolution. In both cases, $B_\t{rf}$ is estimated from a measurement of a component of the collective spin (in the rotating frame), read out, e.g., via the scheme of \citeref{Wasilewski2010}, which resembles a standard Ramsey-interferometry \cite{Wineland1992,Wineland1994,Bollinger1996} measurement. We show that in scenario (b), for one axis-twisted spin-squeezed states, a superclassical scaling $1/N^{5/4}$ of the mean-squared error can be maintained, thus demonstrating that a scaling quantum advantage is possible with feasible states and measurements. At the same time, we find that in the case of (a) the quantum advantage is limited by a constant, which matches the bound for parallel dephasing \cite{Huelga1997,Escher2011}. As a consequence, for an atomic ensemble size and parameters matching the experiment \citeref{Wasilewski2010}, (b) may considerably outperform (a).

As an aside, we note that when the true value of the estimated parameter is zero, the bound of \eqnref{eq.prevbound} vanishes. This does not mean that the precision is unbounded, but indicates that the bound gives no information in such a limit. One may then speculate whether the scaling can be further improved if $\omega$ can be made arbitrarily small in an adaptive manner. We discuss this issue in \appref{app.omega_zero}.

\section{Computing precision for specific states and measurements}

To obtain results for the precision achievable within the above scenarios, we make use of error propagation
and apply it for adequate choices of squeezed states and collective-spin-observable measurements.
Generally, when a parameter $\phi$ is estimated based on the average
of measuring an observable $\hat{O}$, and when the prior knowledge of $\phi$ is sufficiently tight, fluctuations in the estimate can be linearly related to the fluctuations in $\hat{O}$. Thus, for a system in a state $\rho$, in such a local estimation regime the mean-squared error of the estimate may be quantified as
\begin{equation}
\Delta^2\phi = \frac{(\Delta^2 \hat{O})_\rho}{|\partial \langle \hat{O} \rangle_\rho/\partial\phi|^2} = \frac{\langle \hat{O}^2 \rangle_\rho - \langle \hat{O} \rangle_\rho^2}{|\partial \langle \hat{O} \rangle_\rho/\partial\phi|^2}.
\label{eq.errprop}
\end{equation}
If the measurement is repeated, $\Delta^2\phi$ will additionally decrease inverse proportionally to the number of repetitions $\nu$, which also ensures the above local regime as $\nu\!\to\!\infty$ and thus that \eqnref{eq.errprop} always holds. Here, we are interested in frequency estimation over a total time $T$ with a single-round duration $t$, such that $\nu\!=\!T/t$. We therefore write the overall mean-squared error of the $\omega$ estimate as \cite{Giovannetti2006}
\begin{equation}
\label{eq.freqerrprop}
\Delta^2\omega \,T = t \, \frac{(\Delta^2\hat{O})_t}{|\partial \langle \hat{O} \rangle_t/\partial\omega|^2},
\end{equation}
which, like \eqnref{eq.SQLlike}, is valid in the $T\!\gg\!t$ regime, i.e, for sufficiently large $T$. The expectation values in \eqnref{eq.freqerrprop} can be evaluated by computing the expectation value of either the static operator in the time-evolved state or the time-evolved operator in the input state (analogously to the usual Schr\"odinger and Heisenberg pictures for unitary dynamics). Specifically, in terms of the Kraus representation of the evolution, one has
\begin{equation}
\langle \hat{O} \rangle_t = \sum_\mathbf{s} \t{Tr} \left[ \hat{O} K_\mathbf{s}(t) \rho_0 K_\mathbf{s}^\dagger(t) \right] = \sum_\mathbf{s} \t{Tr} \left[ K_\mathbf{s}^\dagger(t) \hat{O} K_\mathbf{s}(t) \rho_0 \right] ,
\end{equation}
where $\hat{O}$ is the time-independent observable, $\rho_0$ is the input state, and $K_\mathbf{s}$ are the Kraus operators of the global channel. For independent channels acting on each qubit, $K_\mathbf{s} = K_{s_1} \otimes \cdots \otimes K_{s_N}$, where the $K_{s_i}$ are the Kraus operators acting on the $i$th qubit.

In subsequent sections, we determine the precisions attainable under our model described by \eqnref{eq.master} for specific input states and measurements. 
The model has four Kraus operators, which have the form
\begin{equation}
\label{eq.krausops}
\begin{split}
K_1 & = a_1 \sh_y , \quad  K_3 = a_3 \sh_z - \ii b_3 \mathbbm{1} , \\
K_2 & = a_2 \sh_x , \quad  K_4 = a_4 \sh_z - \ii b_4 \mathbbm{1} .
\end{split}
\end{equation}
Here, the coefficients $a_i$, $b_i$ are real and depend on the frequency $\omega$, the noise parameters $\gamma$, $\alpha_x$, $\alpha_y$, $\alpha_z$, and the time $t$ (see \appref{app.mapcoeffs}).
However, to simplify notation we suppress these dependences. Because of trace preservation, $\sum_s\! K_s^\dagger K_s \!=\! \mathbbm{1}$, the coefficients must satisfy
\begin{equation}
\label{eq.coeffsum}
a_1^2 + a_2^2 + a_3^2 + a_4^2 + b_3^2 + b_4^2 = 1 .
\end{equation}
For later calculations, it is useful to compute the evolution of
both $\sh_x$ and $\sh_y$ under the Kraus map. For $\sh_x$, we have
\begin{equation}
\begin{split}
K_1^\dagger \sh_x K_1 & = - a_1^2 \sh_x , \quad K_2^\dagger \sh_x K_2  = a_2^2 \sh_x , \\
K_3^\dagger \sh_x K_3 & = (-a_3^2+b_3^2) \sh_x + 2a_3b_3 \sh_x , \\
K_4^\dagger \sh_x K_4 & = (-a_4^2+b_4^2) \sh_x + 2a_4b_4 \sh_x.
\end{split}
\end{equation}
Using \eqnref{eq.coeffsum}, the evolution under the channel can then be written as (Pauli operators with no explicit time dependence are time independent)
\begin{equation}
\label{eq.sigmaxevol}
\sh_x(t) = \sum_s K_s^\dagger \sh_x K_s = \xi_x \sh_x + \chi_x \sh_y ,
\end{equation}
where the coefficients $\xi_x \!=\! 1 - 2(a_1^2 + a_3^2 + a_4^2)$, $\chi_x \!=\! 2(a_3b_3 + a_4b_4)$ are again real and
encode the full dependence of the evolved operator on time, frequency, and the noise parameters. They are given in \appref{app.mapcoeffs}. Similarly, one obtains
\begin{equation}
\label{eq.sigmayevol}
\sh_y(t) = \sum_s K_s^\dagger \sh_y K_s = \xi_y \sh_y + \chi_y \sh_x ,
\end{equation}
with $\xi_y \!=\! 1 - 2(a_2^2 + a_3^2 + a_4^2)$, $\chi_y \!=\! -2(a_3b_3 + a_4b_4)$.

\section{Beating the SQL with realistic states and measurements }

Several experiments have demonstrated superclassical sensitivity of magnetometry with atomic ensembles by squeezing the collective atomic spin \cite{Wasilewski2010,Koschorreck2010,Sewell2012,Ockeloen2013,Sheng2013,Lucivero2014,Muessel2014}. Considering the perpendicular model noise, we now show that spin-squeezed states and Ramsey-type measurements together with interrogation-time optimisation are sufficient not only to reach precisions unattainable by classical protocols but also to maintain superclassical precision scaling with the particle number.

\subsection{Collective spin}

Ramsey interferometry performed on a collection of spin-$1/2$ particles (qubits) effectively corresponds
to collective spin measurements \cite{Wineland1992,Wineland1994,Bollinger1996}.
Here, we consider the components of collective spin along $x$ and $y$,
\begin{equation}
\hat{J}_x = \frac{1}{2} \sum_k \sh_x^{(k)}, \quad \hat{J}_y = \frac{1}{2} \sum_k \sh_y^{(k)},
\end{equation}
which specify the observables measured in scenarios (b) and (a) of \figref{fig.magnetosetup}, respectively.
The evolution of $\hat{J}_x$ under the model of \eqnref{eq.master} follows directly from \eqnref{eq.sigmaxevol},
\begin{equation}
\label{eq.Jxevolv}
\hat{J}_x(t) = \sum_\mathbf{s} K_\mathbf{s}^\dagger \hat{J}_x K_\mathbf{s} = \xi_x \hat{J}_x + \chi_x \hat{J}_y ,
\end{equation}
and similarly for $\hat{J}_y$ using \eqnref{eq.sigmayevol}. The derivatives with respect to the estimated parameter then read
\begin{equation}
\label{eq.Jxexpecevolv}
\frac{\partial \langle \hat{J}_x \rangle_t}{\partial \omega} = \frac{\partial\xi_x}{\partial \omega} \langle \hat{J}_x \rangle_0 +
\frac{\partial\chi_x}{\partial \omega} \langle \hat{J}_y \rangle_0 ,
\end{equation}
and similarly for $\hat{J}_y$ after interchanging $x\!\leftrightarrow\!y$.
We also compute [note that taking the square and evolving do not commute because the evolution is not unitary, i.e, $\hat{J}_x^2(t)\!\not\equiv\!(\hat{J}_x(t))^2$]
\begin{equation}
\label{eq.Jxsqevolv}
\begin{split}
\hat{J}_x^2(t) & = \sum_\mathbf{s} K_\mathbf{s}^\dagger \hat{J}_x^2 K_\mathbf{s}
= \frac{1}{4} \sum_{k,k'} \sum_\mathbf{s} K_\mathbf{s}^\dagger \sh_x^{(k)}\sh_x^{(k')} K_\mathbf{s}  \\
& = \frac{N}{4} + \frac{1}{4}\! \sum_{k\neq k'} \!\sh_x^{(k)}\!(t) \otimes \sh_x^{(k')}\!(t)  \\
& = \frac{N}{4} + \frac{1}{4} \!\left(\!\sum_{k} \sh_x^{(k)}\!(t)\!\right)^2 \!\!- \frac{1}{4} \sum_{k} (\sh_x^{(k)}\!(t))^2 \\
& = \frac{N}{4} + (\hat{J}_x(t))^2 - \frac{1}{4} \sum_{k} (\xi_x \sh_x^{(k)} + \chi_x \sh_y^{(k)})^2 \\
& = \frac{N}{4}(1 - \xi_x^2 - \chi_x^2)+ (\xi_x \hat{J}_x + \chi_x \hat{J}_y)^2,
\end{split}
\end{equation}
so that from Eqs.~\eref{eq.Jxevolv} and \eref{eq.Jxsqevolv} we obtain the variance
\begin{equation}
\label{eq.Jxvarevolv}
\begin{split}
(\Delta^2\hat{J}_x)_t = & \frac{N}{4}(1 - \xi_x^2 - \chi_x^2)
+ \xi_x^2 (\Delta^2 \hat{J}_x)_0 \\
& + \chi_x^2 (\Delta^2 \hat{J}_y)_0 + \xi_x\chi_x (\text{Cov}(\hat{J}_x,\hat{J}_y))_0 ,
\end{split}
\end{equation}
with $\text{Cov}$ denoting the covariance. The variance for $\hat{J}_y$ is
again obtained by just replacing $x\! \leftrightarrow \!y$.

For a specific initial state of the atomic ensemble
with both its expectation values and variances known at $t\!=\!0$,
we can substitute the above expressions into \eqnref{eq.freqerrprop},
in order to quantify the precision attained in
scenarios (a) and (b) of \figref{fig.magnetosetup} for a given interrogation time $t$.

\subsection{One-axis-twisted spin-squeezed states}

There is no unique definition of spin squeezing \cite{Ma2011}, but generally spin-squeezed states are states in which fluctuations of the collective spin component are reduced in a particular direction, when compared to the value they would have in a state with all individual spins aligned, i.e, in a \emph{coherent spin state} (CSS), an eigenstate of the corresponding spin component with maximal eigenvalue. Spin-squeezed states are useful for metrology due to their enhanced sensitivity to any change of the collective spin in the squeezed direction, e.g., caused by precession in a magnetic field.

A number of experiments, in particular \citeref{Wasilewski2010}, employ the so-called two-axis-twisted spin-squeezed states, which can be generated by quantum nondemolition measurement of the collective atomic spin mediated by light. However, here we focus on \emph{one-axis-twisted spin-squeezed states} (OATSSs) because they are amenable to analytical treatment. As two axis-twisted states allow for stronger suppression of the collective spin variance in a particular direction, i.e, stronger squeezing, we expect them to attain precisions at least as good as those derived below for OATSSs. At the same time, quantum advantage with OATSSs in magnetometry has been demonstrated in the experiment of \citeref{Muessel2014} using a Bose-Einstein condensate, and the generation of OATSSs using nitrogen-vacancy centres in diamond has been studied \cite{Bennett2013}.

OATSSs are a particular kind of spin-squeezed states first introduced by Kitagawa and Ueda \cite{Kitagawa1993}.
They can be produced by first preparing atoms in a CSS along one direction, and then applying an evolution with a Hamiltonian quadratic in one of the perpendicular spin components. For example, for spin-$1/2$ particles, one can start from an eigenstate of $\hat J_x$ with eigenvalue $N/2$
(all spins aligned along $x$) and apply an evolution with a Hamiltonian proportional to $\hat{J}_z^2$. This will generate a state with minimum uncertainty at an angle to both the $y$ and $z$ axes, which depends on the strength of the evolution. The state can then be rotated to align the direction of minimum uncertainty with one of the axes.

For scenarios (a) and (b) of \figref{fig.magnetosetup}, we consider two cases where the initial CSS is along either $x$ or $y$, and the collective spin component with minimum uncertainty is $\hat J_y$ or $\hat J_x$, respectively. For scenario (a), the mean values of the collective spin are \cite{Kitagawa1993}
\begin{equation}
\label{eq.oatssexpec}
\left\langle \hat{J}_{x}\right\rangle_0 = \frac{N}{2}\,\cos^{N-1}\frac{\mu}{2},\quad\left\langle \hat{J}_{y}\right\rangle_0 =\left\langle \hat{J}_{z}\right\rangle_0 = 0,
\end{equation}
whereas the variances read
\begin{equation}
\begin{split}
\label{eq.oatssvar}
(\Delta^{2}\hat{J}_{x})_0 & = \frac{N}{4}\left[N\left(1-\cos^{2(N-1)}\frac{\mu}{2}\right)-\frac{1}{2}\left(N-1\right)A\right]\\
(\Delta^{2}\hat{J}_{y})_0 & = \frac{N}{4}\left[1+\frac{1}{4}\left(N-1\right)\left[A-\sqrt{A^{2}+B^{2}}\right]\right]\\
(\Delta^{2}\hat{J}_{z})_0 & = \frac{N}{4}\left[1+\frac{1}{4}\left(N-1\right)\left[A+\sqrt{A^{2}+B^{2}}\right]\right],
\end{split}
\end{equation}
with $\mu$ being the squeezing parameter, $A\!=\!1-\cos^{N-2}\mu$, and $B\!=\!4\sin\frac{\mu}{2}\cos^{N-2}\frac{\mu}{2}$. We note that the covariance $(\text{Cov}(\hat{J}_x,\hat{J}_y))_0\!=\! 0$ vanishes for this state. The equivalents of Eqs.~\eref{eq.oatssexpec} and \eref{eq.oatssvar} for scenario (b) are obtained by interchanging $x\!\leftrightarrow\!y$.

\subsection{Mean-squared-error scaling under transversal noise}
\label{sub.varscaling}

The mean-squared errors of estimation, which are achieved in scenarios (a) and (b) of \figref{fig.magnetosetup}, can be calculated by using Eqs.~\eref{eq.oatssexpec}, \eref{eq.oatssvar}, \eref{eq.Jxexpecevolv}, \eref{eq.Jxvarevolv} (and the equivalents for $\hat{J}_y$) and substituting into \eqnref{eq.freqerrprop}. The best precision is then obtained by optimising the evolution time $t$ and the squeezing $\mu$ for each $N$. The general expressions are rather involved, and we have been able to obtain their minima only numerically. However, any explicit choice of $t(N)$ and $\mu(N)$ provides a precision that is guaranteed to be attainable.

Specifically, for scenario (b) a choice that appears to be nearly optimal is $\mu\!=\!(\gamma/\omega)^{1/4}(N/4)^{-4/5}$ and $t\!=\!(\gamma\omega)^{-1/2}N^{-1/8}$ \footnote{Although $\omega$ is not perfectly known in advance, this choice is not a problem. Replacing $\omega \rightarrow \beta\omega$ in $\mu$, $t$ has the effect $\omega \rightarrow \omega/\beta$ in \eqnref{eq.squeezenonzeroperpscaling}. Similarly, $\gamma \rightarrow \beta\gamma$ in $\mu$, $t$ leads to $\omega \rightarrow \omega/\beta$ in \eqnref{eq.squeezenonzeroperpscaling}. The scaling remains unchanged.}. For this choice, we expand \eqnref{eq.freqerrprop} in $1/N$ to find the expression for the asymptotic mean-squared error
\begin{equation}
\label{eq.squeezenonzeroperpscaling}
\Delta^2 \omega_\t{(b)}T \;\underset{N\to\infty}{=}\;\frac{2\omega}{3}\frac{1}{N^{5/4}} .
\end{equation}
Since the scaling is better than the $1/N$ of the SQL, this demonstrates that superclassical precision scaling is indeed possible with spin-squeezed states and Ramsey-type measurements in the presence of transversal noise.

The possibility for a large quantum enhancement depends strongly on the geometry. We can see this by comparing with scenario (a). There, we find that
\begin{equation}
\label{eq.squeezenonzeroperpscalingA}
\Delta^2 \omega_\t{(a)} T \;\underset{N\to\infty}{=}\; \frac{2\gamma}{N} .
\end{equation}
which coincides with the best achievable precision for the parallel-noise setting \cite{Escher2011} constrained by \eqnref{eq.cZ}. As discussed in \appref{app.scenarioabound}, we find an analytical proof of \eqnref{eq.squeezenonzeroperpscalingA} in the limit $\omega\!\to\!0$, and strong numerical evidence for arbitrary $\omega$, which indicates very clearly that no better precision can be achieved. The value is attainable by any choice of $\mu\!\propto\!1/N^{s/(s+1)}$ and $t\!\propto\!1/N^s$ with $s\!>\!1$. Thus, for this geometry, under transversal noise only SQL-like scaling is possible and the quantum enhancement over classical, nonentangled strategies is bounded, while for scenario (b), the quantum enhancement is unbounded with increasing $N$.

One can understand intuitively why scenario (b) is more robust than (a) from the pictures of the collective spin in \figref{fig.magnetosetup}. In both cases the dephasing is directed along $x$ and so causes random rotations around the $x$ axis. The effect of such rotations on the state in (a) is to smear the squeezed state into a more circular distribution at the pole. This reduces the sensitivity of a $\hat{J}_y$ measurement to a small rotation around the z axis, which is the signal we want to detect. In scenario (b), the effect of $x$ rotations is to smear the state along the equator. However, this does not affect the sensitivity of the $\hat{J}_x$ measurement to a $z$ rotation as strongly. Using the same picture, one can also understand why there is a finite optimal value of the squeezing parameter $\mu$. When $\mu$ becomes large, the ellipse starts to stretch around the ball. Referring to scenario (b), a $z$ rotation will then increase the projection of the state onto the $x$ axis and hence the $\hat{J}_x$ measurement loses sensitivity.

\begin{figure}
\begin{center}
\includegraphics[width=0.99\linewidth]{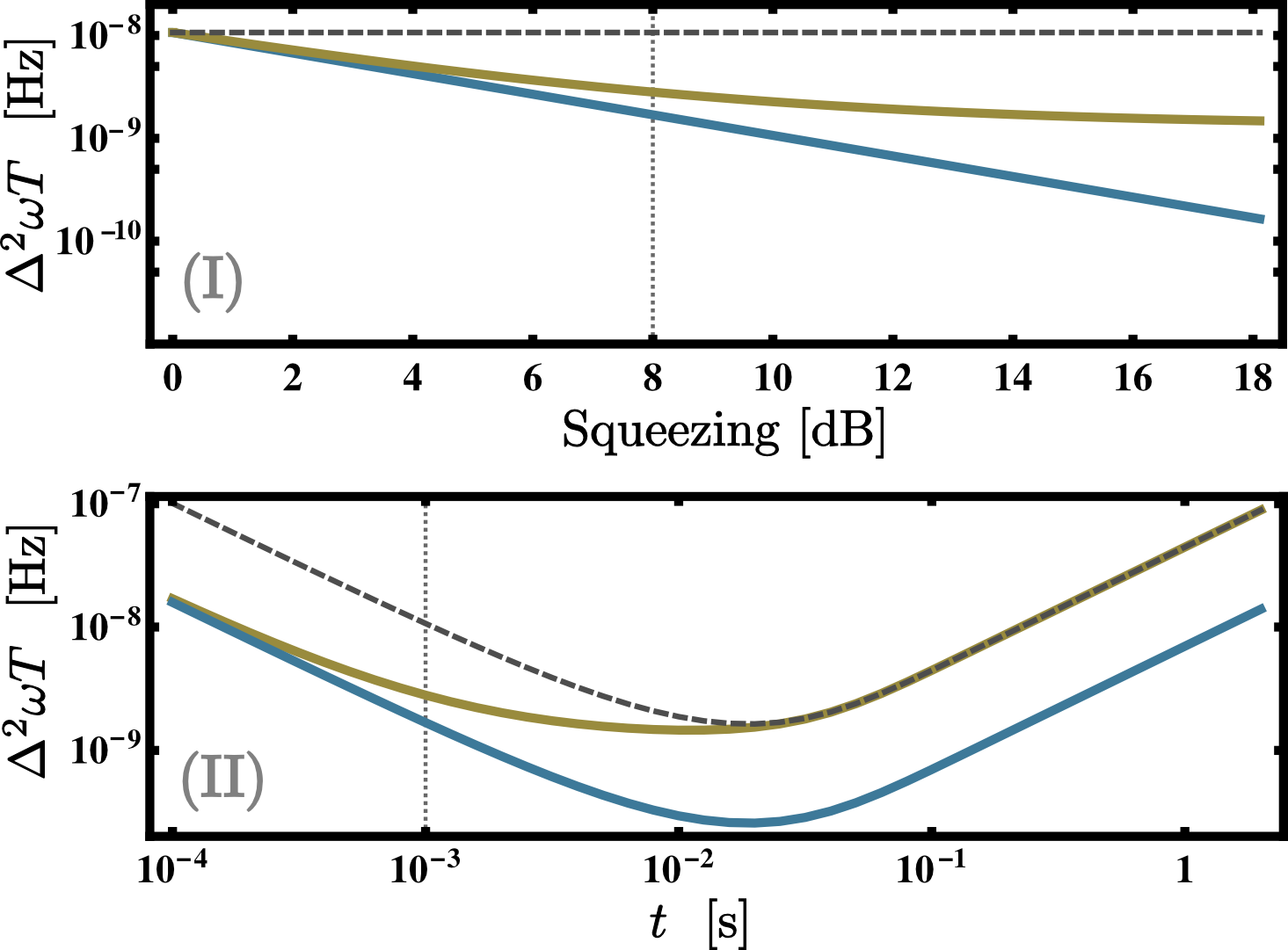}
\end{center}
\caption{Mean-squared error of estimation as a function of (I) squeezing and (II) interrogation time, for parameters corresponding to \cite{Wasilewski2010}, $N\!=\!10^{11}$, $\gamma \!=\! 67\,\t{Hz}$, $\omega \!=\! 3.6\times 10^{-3}\,\t{Hz}$. In (I) the interrogation time is fixed at $t\!=\!1\,\t{ms}$ while in (II) the squeezing is fixed at $-8\,\t{dB}$ (indicated by the dotted lines, these values correspond to the numbers discussed in the text). The results for scenario (a) (\emph{yellow line}), scenario (b) (\emph{blue line}), and for CSS (\emph{dashed gray line}) are shown.}
\label{fig.var_vs_tmu}
\end{figure}

The difference between the two geometries shows up only for quantum strategies, that is, when squeezed states are employed. If the initial states are not squeezed but are simply CSS states along $x$ for scenario (a) or $y$ for scenario (b), then the precision takes the same form in both cases:
\begin{equation}
\label{eq.CSSscaling}
\Delta^2 \omega_\t{\tiny CSS}T=\!\frac{2t\gamma^{2}\Gamma^{2}\bar{\Gamma}^{4}\ee^{-\gamma t\bar{\Gamma}}\!\left[1\!-\!\cosh\!\left(\gamma t\bar{\Gamma}\right)\!+\!2\ee^{\gamma t}\!\left(\frac{\bar{\Gamma}}{\Gamma}\right)^{2}\right]}
{\left[2\left(1\!-\!\ee^{-\gamma t\bar{\Gamma}}\right)-\gamma t\Gamma^{2}\bar{\Gamma}\left(1\!+\!\ee^{-\gamma t\bar{\Gamma}}\right)\right]^{2}}
\frac{1}{N},
\end{equation}
where $\Gamma\!=\!2\omega/\gamma$ and $\bar\Gamma\!=\!\sqrt{1-\Gamma^2}$. Thus, we can benchmark the quantum enhancement in either scenario against this classical value. In particular, we consider the numbers from \citeref{Wasilewski2010}. In this experiment, $N \!\approx\! 10^{11}$, $T_2 \!\approx\! 30\,\t{ms}$, $\kappa \!\approx\! 10^{10}\,(\t{Ts})^{-1}$, and $B_\t{rf}\!\approx\!36\,\t{fT}$, which gives $\gamma\!=\!2/T_2\!\approx\!67\,\t{Hz}$ and $\omega\!=\!2\kappa B_\t{rf} \approx 3.6\times 10^{-3}\,\t{Hz}$, and the measurement time is $t\!\approx\!1\,\t{ms}$. The experiment is not performed with OATSSs, but we compute the quantum enhancements that OATSSs would provide. We insert the numbers in the full expressions [from \eqnref{eq.freqerrprop}] for $\Delta^2 \omega_\t{(a)}T$ and $\Delta^2 \omega_\t{(b)}T$ and vary the squeezing, which we quantify in dB \footnote{We measure the squeezing by the CSS-relative squeezing parameter $\xi_\t{rel}$ introduced in \cite{Wineland1992,Wineland1994} and use Eqs.~(112) and (139) of \cite{Ma2011}.}; see \figref{fig.var_vs_tmu}(I). The precision in scenario (a) saturates with increasing squeezing, and the best quantum enhancement attainable is $\Delta^2 \omega_\t{\tiny CSS} / \Delta^2 \omega_\t{(a)}\!\approx\!8$. In scenario (b), on the other hand, the enhancement can reach $\Delta^2 \omega_\t{\tiny CSS} / \Delta^2 \omega_\t{(b)} \!\approx\! 2\times 10^7$, corresponding to a 4500-fold improvement in precision. In this case, the precision does not saturate but is limited by the maximal squeezing attainable by the OATSS. This underlines the advantage offered by geometry. However, these maximal enhancements require rather prohibitive squeezings of $ -18\,\t{dB}$ and $-73\,\t{dB}$, respectively. If we restrict the squeezing to at most $-8\,\t{dB}$ as discussed in the outlook of \citeref{Wasilewski2010}, then scenario (b) provides an enhancement of $\Delta^2 \omega_\t{\tiny CSS} / \Delta^2 \omega_\t{(b)} \!\approx\! 6.4$, corresponding to a factor of 2.5 in precision, while scenario (a) for the same numbers gives $\Delta^2 \omega_\t{\tiny CSS} / \Delta^2 \omega_\t{(a)} \!\approx\! 3.8$ corresponding to a factor of 1.95. However, the behaviour in the two scenarios for varying interrogation time is very different, as explicitly shown in \figref{fig.var_vs_tmu}(II). The performance is similar at short times, but when the interrogation time is increased, the quantum advantage in scenario (a) is lost, while the advantage in scenario (b) is maintained. This is a nice feature of the optimized geometry, as in practice experimental constraints may impose a lower limit on $t$. As seen from \figref{fig.var_vs_tmu}(II), at a squeezing of $-8\,\t{dB}$, the best precision in scenario (a) (optimising $t$) is not significantly below the best precision attainable with a CSS, while the advantage of scenario (b) importantly remains even when experimental constraints do not allow for arbitrarily small $t$.

We note that, as for \eqnref{eq.prevbound}, the error \eref{eq.squeezenonzeroperpscaling} vanishes as $\omega \!\rightarrow\! 0$. We refer the reader to \appref{app.omega_zero} for a discussion of this limit.

\subsection{Non-transversal noise sources}
\label{sub.non_trans_noise}

In a realistic implementation, in addition to the dominant transversal noise, other sources of decoherence will be present. For example, in some setups different from \citeref{Wasilewski2010},
e.g., spin-exchange relaxation-free magnetometers \cite{Ledbetter2008,Griffith2010}, spin relaxation cannot be neglected. Typically \cite{Kominis2008,Liu2015}, this is modelled as uncorrelated depolarising noise [$\alpha_x\!=\!\alpha_y\!=\!\alpha_z\!=\!1/3$ in \eqnref{lindblad}] with a strength dictated by the spin-relaxation time $T_1$ (see \appref{app.master}). As opposed to the directional dephasing caused by spin decoherence, such noise is isotropic and always yields a parallel-noise component independently of the geometry. In the case of \citeref{Wasilewski2010}, any spatial misalignment between $B$, $B_\t{rf}$, and the direction of squeezing, temporal mismatch between the $B_\t{rf}$ rotations, and the Larmor frequency of $B$, or violation of the condition $B_\t{rf}\!\gg\!B$, may be phenomenologically included into such a depolarisation model.

\begin{figure} [!t]
\centering
\includegraphics[width=0.995\columnwidth]{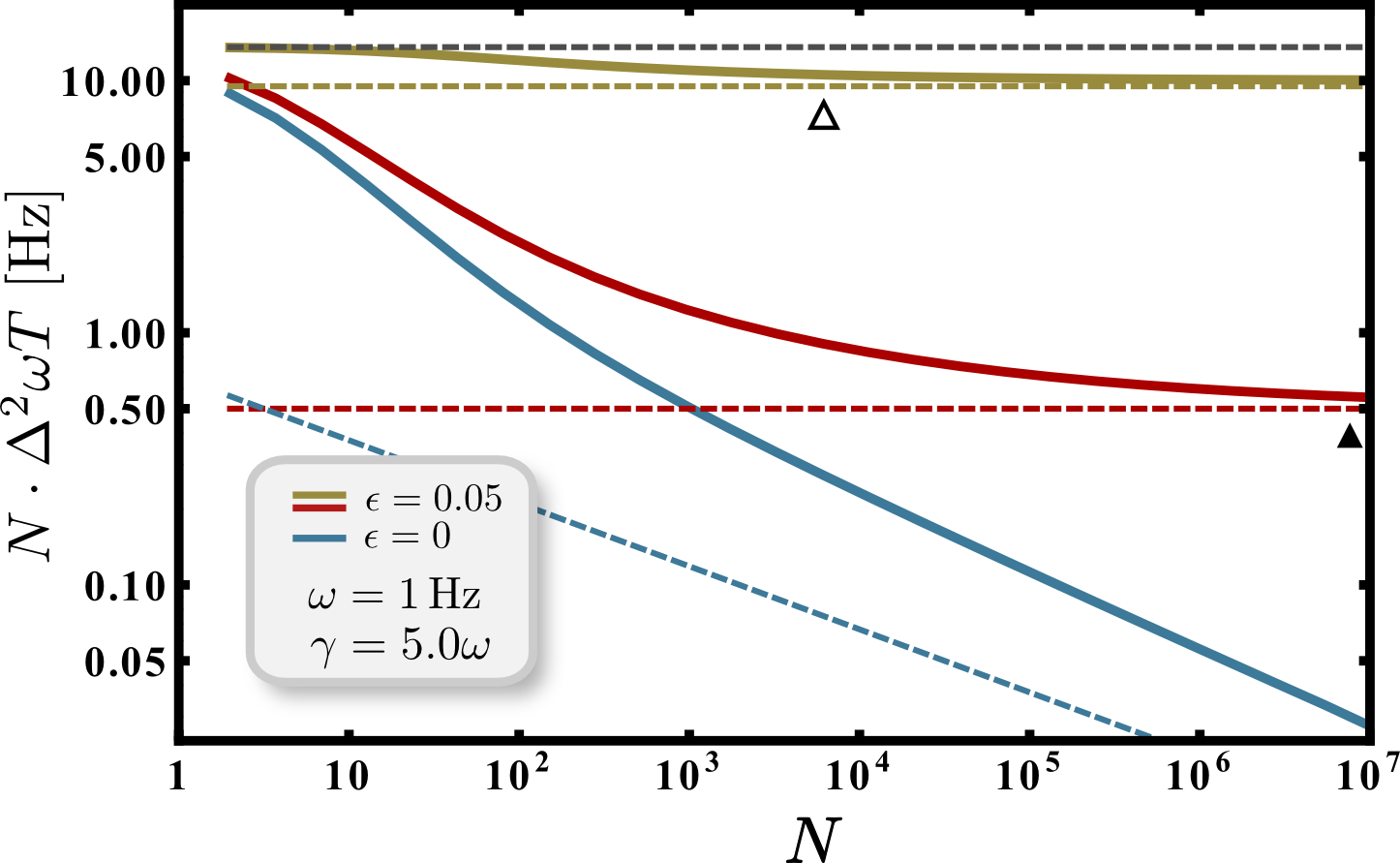}
\caption{The mean-squared error of estimation rescaled by the particle number such that SQL-like scaling is horizontal. The curves for scenario (a) (\emph{yellow line}) and (b) (\emph{red line}) with 5\% of parallel noise are shown, and for scenario (b) under purely transversal noise (\emph{blue line}), along with their asymptotes (\emph{dashed line}), respectively $2\gamma(1-\epsilon)N$ from \eqnref{eq.squeezenonzeroperpscalingA}, $2\gamma\epsilon$ from the parallel noise component, and $2\omega/3N^{5/4}$ from \eqnref{eq.squeezenonzeroperpscaling}. We also show the performance of a CSS without squeezing (\emph{grey dashed line}) as described by \eqnref{eq.CSSscaling}. The locations where the OATSS strategies reach 90\% of their asymptotic gain over this CSS strategy are indicated for scenarios (a) (\emph{open triangle}) and (b) (\emph{closed triangle}).}
\label{fig.deviations}
\end{figure}

We assess the effect of such additional noise sources by considering a deviation from perfect transversality. In particular, we take a small component of dephasing directed along the $z$ axis, such that $\alpha_{x}\!=\!1-\epsilon$ and $\alpha_z\!=\! \epsilon$. As discussed in \citeref{Chaves2013}, once any such parallel-dephasing contribution is present, the asymptotic scaling must return to its SQL-like behaviour; that is,
\begin{equation}
\label{eq.xzbound}
\Delta^2 \omega T  \ge  \frac{c_{xz}(\gamma,\epsilon)}{N}
\end{equation}
with $c_{xz}(\gamma,\epsilon)$ lower-bounded by the minimum of \eqnref{eq.cZ}, i.e, $c_z(\epsilon\gamma,\omega,t)\!\ge\! 2\epsilon\gamma$. For instance, for the depolarisation-based spin-relaxation model, $c_{xz}(\gamma,\epsilon)\!\ge\!8/(3T_1)$ (see \appref{app.master}). We illustrate the resulting crossover behaviour in \figref{fig.deviations}.

Although the asymptotic scaling is now again SQL-like, geometry can strongly influence the achievable quantum gain and the effective $N$ at which the crossover to SQL-like scaling happens. In \figref{fig.deviations} we show the mean-squared-error scalings attained in scenarios (a) and (b) using, respectively, OATSSs along $x$ and $y$ and measurements of $\hat{J}_y$ and $\hat{J}_x$ and compare them to a strategy without entanglement, simply using a CSS along $x$ and measurement of $\hat{J}_y$ corresponding to the nonentangled strategy implemented in \citeref{Wasilewski2010}. We see that while the strategy in (b) can saturate the bound $2\gamma\epsilon/N$, the strategy in (a) only reaches $2\gamma(1-\epsilon)/N$ as imposed by \eqnref{eq.squeezenonzeroperpscalingA}. Thus, the mean-squared error of geometry (b) is a factor $\epsilon/(1-\epsilon)$ lower than (a), which may be significant when the noise is dominantly transversal. Furthermore, superclassical scaling persists over a larger range of $N$ in geometry (b). In the figure, the locations where the OATSS strategies for (a) and (b) reach 90\% of their asymptotic gain over the nonentangled CSS strategy are indicated. Clearly, the crossover happens at much larger $N$ in scenario (b). As $\epsilon\! \rightarrow\! 0$, the crossover must go to infinity. To get an idea of the behaviour, we can take the $N$ at which the asymptotic bound $2\epsilon\gamma/N$ crosses the asymptote \eref{eq.squeezenonzeroperpscaling} for perfectly transversal noise. This intersection scales as $(\omega/\gamma)^4/\epsilon^4$. Thus, significant gain in precision by squeezing is attained over a larger range of $N$ if the geometry is chosen correctly.

\section{Conclusion and outlook}

For quantum metrology to be relevant in practical situations, it is important that good performance can be attained under realistic noise with states and measurements that are amenable to implementation in the laboratory. While recent results have shown that for many noise types precision scaling can only improve over the classical limit by a constant, here we demonstrate that under transversal dephasing, superclassical scaling can be preserved with experimentally accessible states and measurements, and we argue that this noise model is relevant to recent atomic magnetometry experiments. We show that the choice of geometry is important for the attainable quantum improvement both asymptotically and for parameter settings corresponding to recent experiments. Furthermore, we assess the robustness of the model to other nontransversal sources of noise and find that quantum enhancement could still be achieved for atomic ensembles of macroscopic size with an adequate choice of geometry.

Our results give a clear message that quantum-enhanced metrology maintains its relevance even in the presence of noise, and we hope that they will encourage the search for other practically motivated scenarios where quantum strategies provide an advantage. For instance, it has been suggested that the transversal-noise model applies also to nitrogen-vacancy centres in diamonds \cite{Kessler2014}, and that one-axis-twisted spin-squeezed states could be prepared in such systems \cite{Bennett2013}. Very recently, a noise-robust magnetometry scheme employing SQUID junctions has been proposed \cite{Herrera2014}. Finally, in the Appendices, we speculate about the potential of adaptive techniques that bias the estimated parameter towards the zero value, for which our current precision bounds fail. We expect that the question of what happens in this limit could be consistently resolved by employing Bayesian techniques \cite{Kolodynski2014,Macieszczak2014,Jarzyna2014}, which account explicitly for the prior knowledge about the estimated parameter.

\section{Acknowledgements}

We are grateful to Antonio Ac\'{i}n, Stefan Christensen, Rafa\l{} Demkowicz-Dobrza\'{n}ski, Marcin Jarzyna and Wojciech Wasilewski for valuable feedback and discussions. J.~B.~B. acknowledges support from the Swiss National Science Foundation (QSIT director's reserve), SEFRI (COST action MP1006), and FP7 IP project SIQS. R. C. acknowledges support from the Excellence Initiative of the German Federal and State Governments (Grant ZUK 43 and 81), the Research Innovation Fund from the University of Freiburg, and the U.S.~Army Research Office under Contracts No.~W911NF-14-1-0098 and No.~W911NF-14-1-0133 (Quantum Characterisation, Verification, and Validation).
J.~K.~acknowledges support from the FP7 IP project SIQS co-financed by the Polish Ministry of Science and Higher Education, START scholarship granted by Foundation for Polish Science, the ERC Consolidator Grant QITBOX, as well as the European Union’s Horizon 2020 research and innovation programme under the Marie Sklodowska-Curie grant agreement 
No.~655161 (Q-METAPP).

\appendix

\section{Atomic master equation}
\label{app.master}

Following \citeref{Wasilewski2010}, we assume correlated effects in the atomic ensemble to play a role only at time scales longer than the interrogation time, i.e, the ensemble relaxation and decoherence times $T_1^{*},T_2^{*}\!\gg\!t$ \cite{Wang2005} (in \citeref{Wasilewski2010}, $t \approx\!\!1\t{ms}$). We may then describe the dynamics by a master equation where each atom evolves independently
\begin{equation}
\frac{\partial\rho^{N}(t)}{\partial t}\;=\;\sum_{n=1}^{N}\,\mathcal{D}\!\!\left[\rho^{N}(t)\right]
\end{equation}
where $\mathcal{D}$ generates the evolution for a single atom (assumed to be the same for all atoms) and reads
\begin{eqnarray}
\mathcal{D}\!\left[\rho(t)\right] & = & -\ii\frac{\Omega_B}{2}\left[\hat{\sigma}_{x},\rho(t)\right]-\ii\frac{\omega}{2}\left[\hat{\sigma}_{yz}(t),\rho(t)\right] \label{eq.gen_sing_part_H}\\
 &  & +\frac{1}{T_{2}}\left(\hat{\sigma}_{x}\rho(t)\hat{\sigma}_{x}-\rho(t)\right) \label{eq.gen_sing_part_T2}\\
 &  & +\frac{1}{T_{1}}\left(\frac{1}{3}\sum_{i=1}^{3}\hat{\sigma}_{i}\,\rho(t)\hat{\sigma}_{i}-\rho(t)\right). \label{eq.gen_sing_part_T1}
\end{eqnarray}

The terms \eref{eq.gen_sing_part_H} represent the Hamiltonian part of the dynamics:~$\Omega_B\!=\!2\kappa B$ is the Larmor frequency of the strong field $B$ (see \figref{fig.magnetosetup}), whereas $\omega\!=\!2\kappa B_\t{rf}$ is the frequency corresponding to the weak field $B_\t{rf}$ being sensed. Having \citeref{Wasilewski2010} in mind, we allow $B_\t{rf}$ to rotate in the $yz$ plane $\hat\sigma_{yz}(t)\!=\!\cos\theta_t \,\hat\sigma_z-\sin\theta_t \,\hat\sigma_y$.

The term listed in \eqnref{eq.gen_sing_part_T2} represents dephasing noise. It is a consequence of $B$-field fluctuations independently affecting each atom (e.g., arising from the atomic motion and/or local field inhomogeneities). Equivalently, it may be derived by considering an average-description of a noiseless evolution but with $\Omega_B$ fluctuating according to a Gaussian distribution with variance $4/(T_2t)$ (note that the variance diverges in the $t\!\to\!0$ limit manifesting the Markovianity of the noise). Such fluctuations define the \emph{spin-decoherence time} $T_2$ \cite{Wang2005} and constitute the transversal \cite{Chaves2013} noise which is the main focus of this paper.

The term \eref{eq.gen_sing_part_T1} represents the spin-relaxation process occurring predominantly due to spin-destruction or spin-exchange atomic collisions \cite{Kominis2008} (the latter may be eliminated in the accordingly called spin-exchange relaxation-free magnetometers \cite{Ledbetter2008,Griffith2010}). Note that the term \eref{eq.gen_sing_part_T1} effectively yields a depolarising channel [as defined in \eqnref{lindblad}] with strength parametrised by the \emph{spin-relaxation time} $T_1$ \cite{Wang2005,Liu2015}.

By moving to the rotating frame precessing with the Larmor frequency, in which $\rho_\t{RF}(t)\!=\!\ee^{\frac{\ii\Omega_Bt}{2}\hat\sigma_x}\rho(t)\,\ee^{-\frac{\ii\Omega_Bt}{2}\hat\sigma_x}$,
and by using that $\ee^{\frac{\ii\theta}{2}\hat{\sigma}_{x}}\hat{\sigma}_{\nicefrac{y}{z}}\ee^{-\frac{\ii\theta}{2}\hat{\sigma}_{x}}\!=\!\cos\theta\,\hat{\sigma}_{\nicefrac{y}{z}}\,\mp\,\sin\theta\,\hat{\sigma}_{\nicefrac{y}{z}}$, we obtain the single-atom master equation in the RF as
\begin{eqnarray}
\frac{\partial\rho_\t{RF}(t)}{\partial t} & = & -\ii\frac{\omega}{2}\left[
\sin\delta\theta_t\,\hat{\sigma}_{y}+\cos\delta\theta_t\,\hat{\sigma}_{z}
,\rho_\t{RF}(t)\right] \nonumber\\
 &  & +\frac{1}{T_{2}}\left(\hat{\sigma}_{x}\rho_\t{RF}(t)\hat{\sigma}_{x}-\rho_\t{RF}(t)\right)\nonumber \\
 &  & +\frac{1}{T_{1}}\left(\frac{1}{3}\sum_{i=1}^{3}\hat{\sigma}_{i}\,\rho_\t{RF}(t)\hat{\sigma}_{i}-\rho_\t{RF}(t)\right).
\label{eq.gen_sing_part_RF}
\end{eqnarray}
with $\delta\theta_t\!=\!\frac{\Omega_Bt}{2}-\theta_{t}$. As a consequence, if the Larmor frequency is chosen to exactly match the rotation of $B_\t{rf}$, i.e, $\theta_t\!=\!\Omega_Bt/2$, and if we consider the limit $T_1\!\to\!\infty$ in which the spin relaxation is completely ignored (as in \cite{Wasilewski2010}), then we indeed recover \eqnref{eq.master_magnetometry} in the main text.

Note that \eqnref{eq.gen_sing_part_RF} indicates that by \emph{not} exactly matching $\theta_t$ with the Larmor frequency we nonetheless preserve the required transversal geometry between the dephasing noise \eqnref{eq.gen_sing_part_T2} and the $\omega$-encoding part, so that the perpendicular-noise model of \eqnref{lindblad} introduced in \citeref{Chaves2013}, in principle, still applies. However, let us emphasize that for our analysis of squeezed states to be valid, the geometry of squeezing must always be adjusted for a particular choice of $B_\t{rf}$ direction in the RF, e.g., depicted in \figref{fig.magnetosetup}
for $B_\t{rf}$ chosen to be stationary and pointing along $z$ in the RF. In an experiment, we expect there to be some mismatch between the $B_\t{rf}$-rotation frequency and $\Omega_B$, which would constitute an extra source of \emph{global} (correlated across particles) dephasing noise along $x$, i.e, decreasing the ensemble spin-decoherence time $T_2^{*}$ neglected in \citeref{Wasilewski2010}. In contrast, any such fluctuations felt locally by atoms would just lower $T_2$ in \eqnref{eq.gen_sing_part_T2}, which would thus not impair the robustness of the setup.

On the other hand, any geometrical misalignment of the fields or any instabilities of $\theta_t$ can always be modelled in \eqnref{eq.gen_sing_part_RF} by lowering the effective spin-relaxation time $T_1$. Importantly, the spin-relaxation process may be interpreted as a nontransversal-noise source, which we study in \secref{sub.non_trans_noise}. Similarly to the parallel dephasing case [$\alpha_z\!=\!1$ in \eqnref{lindblad}], which is known to asymptotically restrict the precision to $2\gamma/N$ \cite{Escher2011}, it has been shown \cite{Knysh2014} that depolarising noise [$\alpha_x\!=\!\alpha_y\!=\!\alpha_z\!=\!1/3$ in \eqnref{lindblad}] yields a SQL-like asymptotic precision $(4\gamma/3)/N$, thus giving just a $2/3$-factor improvement as compared to the pure parallel-noise model.

A finite spin-relaxation time must thus impose an asymptotic SQL-like behaviour, which we may estimate by taking $3/2$ of $T_1$ in \eqnref{eq.gen_sing_part_RF} to contribute solely to the parallel-dephasing component. Then, after substituting for $\epsilon\!=\!2T_2/(3T_1\!+\!2T_2)$ and $\gamma\!=\!2(3T_1\!+\!2T_2)/(3T_1T_2)$, we can directly utilise \eqnref{eq.xzbound} to obtain an asymptotic bound with $2\gamma\epsilon\!=\!8/(3T_1)$ dictated by the spin relaxation. Crucially, the analysis of \secref{sub.non_trans_noise} thus shows that in the regime of $T_1\!\gg\!T_2$, the optimal geometry of scenario (b) not only allows the counterbalance of the spin-decoherence effects, but also postpones the inevitable SQL-bounding impact of spin relaxation to much higher $N$ (see the triangular marks in \figref{fig.deviations}).

\section{Kraus operators}
\label{app.mapcoeffs}

The map corresponding to evolution under the master equation \eqnref{eq.master} during time $t$ can be written as a composite map of the form $\mathcal{E}_{\omega}^{\otimes N}$. Following Andersson \textit{et al.}~\cite{andersson2007}, the single-qubit maps are then given by
\begin{equation}
\mathcal{E}_{\omega}\! \left( \rho \right) = \sum_{i,j=0}^3 S_{ij}\, \tilde{\sigma}_{i} \rho \tilde{\sigma}_{j} ,
\label{map}
\end{equation}
where $\tilde{\sigma}_{i}$ are the normalised Pauli operators $\tilde{\sigma}_{i} \!=\! \hat \sigma_{i}/\sqrt{2}$ and $\hat\sigma_{0}$ denotes the identity. All elements of the matrix S are zero, except %
$S_{00}\!=\!A_+\!+\!B_{+}$,
$S_{11}\!=\!A_-\!+\!\frac{\Gamma}{\tilde{\alpha}}B_{-}$,
$S_{22}\!=\!A_-\!-\!\frac{\Gamma}{\tilde{\alpha}} B_{-}$,
$S_{33}\!=\!A_+\!-\!B_{+}$,
$S_{03}\!=\!\ii \frac{\alpha_{-}}{\tilde{\alpha}} B_{-}$,
$S_{03}\!=\!-\ii \frac{\alpha_{-}}{\tilde{\alpha}} B_{-}$.
Where we have defined $\Gamma\!=\!2\omega/\gamma$, $\alpha_{\pm}\!=\!\alpha_{x}\pm\alpha_{y}$, and $\tilde{\alpha}\!=\!\sqrt{\alpha_{-}^{2}-\Gamma^{2}}$, and the coefficients
\begin{align}
A_{\pm} & =\frac{1}{2}\left(1\pm\ee^{-\gamma t\alpha_{+}}\right)\! , \\
B_{\pm} & =\frac{1}{2}\ee^{-\frac{\gamma t}{2}\left(1+\alpha_{z}-\tilde{\alpha}\right)}\!\left(1\pm\ee^{-\gamma t\tilde{\alpha}}\right)\!.
\end{align}

A Kraus representation of the map $\mathcal{E}_{\omega}$ can be obtained by diagonalising the matrix $S$.
Denoting the eigenvalues and normalised eigenvectors of $S$ by $\lambda_i$ and $\mathbf{v}_i$ respectively,
one can find a valid set of Kraus operators for the channel:
\begin{equation}
K_j = \sum_{i=1}^4 \sqrt{|\lambda_i|} \, (\mathbf{v}_i)_j \,\tilde{\sigma}_{j-1}
\end{equation}
with $j\!=\!1,\dots,4$, which gives the set in \eqnref{eq.krausops}. The coefficients in \eqnref{eq.krausops} are rather involved, and we do not explicitly state them here. Instead, we directly give the expressions for $\xi_x$, $\chi_x$, $\xi_y$, and $\chi_y$ of Eqs.~\eref{eq.sigmaxevol} and \eref{eq.sigmayevol}.
For general noise they read
\begin{align}
\xi_x & =\ee^{-\frac{\gamma t}{2}(1+\alpha_{z})}\!\left[\cosh\!\left(\frac{\gamma t}{2}\tilde{\alpha}\right) + \frac{\alpha_{-}}{\tilde{\alpha}}\sinh\!\left(\frac{\gamma t}{2}\tilde{\alpha}\right)\right], \nonumber \\
\chi_y & = - \,\ee^{-\frac{\gamma t}{2}(1+\alpha_{z})}\frac{\Gamma}{\tilde{\alpha}}\sinh\!\left(\frac{\gamma t}{2}\tilde{\alpha}\right) ,
\end{align}
and
\begin{align}
\xi_y & =\ee^{-\frac{\gamma t}{2}(1+\alpha_{z})}\!\left[\cosh\!\left(\frac{\gamma t}{2}\tilde{\alpha}\right) - \frac{\alpha_{-}}{\tilde{\alpha}}\sinh\!\left(\frac{\gamma t}{2}\tilde{\alpha}\right)\right], \nonumber \\
\chi_y & =\ee^{-\frac{\gamma t}{2}(1+\alpha_{z})}\frac{\Gamma}{\tilde{\alpha}}\sinh\!\left(\frac{\gamma t}{2}\tilde{\alpha}\right).
\end{align}
In the case of perfectly transversal noise they further simplify, since $\alpha_x\!=\!1$ implies $\alpha_z\!=\!0$ and $\tilde\alpha\!=\!\sqrt{1-\Gamma^2}$.

\section{Analytical scaling for GHZ states}
\label{app.ghz_states}

Stemming from the error-propagation method [see \eqnref{eq.freqerrprop}] utilised in the main text,
we can also confirm the results of \citeref{Chaves2013} analytically for the \emph{GHZ input states}:
\begin{equation}
\label{eq.GHZ}
\ket{\t{GHZ}} = \frac{1}{\sqrt{2}} ( \ket{0,\ldots,0} + \ket{1,\ldots,1} ),
\end{equation}
by considering the \emph{parity operator} in the $x$ direction:
\begin{equation}
\label{eq.parity}
\hat{P}_x = \otimes_{k=1}^N\, \sh_x^{(k)}\;,
\end{equation}
as the observable being measured.
Similarly to the case of collective spin operators and \eqnref{eq.Jxevolv}, we may utilise
\eqnref{eq.sigmaxevol} to write the form of the parity operator at time $t$ as
\begin{equation}
\label{eq.Pxevol}
\hat{P}_x(t) = \sum_\mathbf{s} K_\mathbf{s}^\dagger \hat{P}_x K_\mathbf{s} = \otimes_{k=1}^N \left( \xi_x \,\sh_x^{(k)} + \chi_x\, \sh_y^{(k)} \right) .
\end{equation}
In the computational basis $\{\ket{0},\ket{1}\}^{\otimes N}$, such an operator just flips all of the qubits, and hence
only the off-diagonal terms contribute when calculating its expectation value for a GHZ state of \eqnref{eq.GHZ}.
Every $\sh_x$ contributes a factor of 1 while $\sh_y$ contributes a factor of $\pm \ii$.
Thus, the expectation value of the measurement becomes
\begin{equation}
{\langle \hat{P}_x \rangle}_{\t{\tiny GHZ},t} = \frac{1}{2} \left[ (\xi_x + \ii \chi_x)^N + (\xi_x -\ii \chi_x)^N \right],
\end{equation}
and, since $\hat P_x^2 \!=\! \mathbbm{1}$, it follows that $\Delta^2 \hat P_x \!=\! 1 \!-\! \langle \hat P_x \rangle^2$.

We compute the mean-squared error of estimation via \eqnref{eq.freqerrprop}, after setting the interrogation time to $t \!=\! (3/\gamma\omega^2N)^{1/3}$, as was found in \citeref{Chaves2013} from numerical analysis. Expanding the corresponding $\Delta^2\omega\,T$ in $1/N$, we find the asymptotic scaling
to read:
\begin{eqnarray}
\label{eq:parity_bound}
\Delta^2 \omega T & \underset{N\to\infty}{=} & g(\gamma,\omega,N) (\gamma\omega^2)^{1/3} \,\frac{1}{N^{5/3}} \nonumber \\
& \geq & \frac{\ee^2 }{3^{1/3} } (\gamma\omega^2)^{1/3}\, \frac{1}{N^{5/3}} ,
\end{eqnarray}
where $g(\gamma,\omega,N)$ represents oscillating terms that are lower-bounded by $\ee^2/3^{1/3}$. The constant prefactor here is larger than the prefactor $3^{2/3}/2$, which was numerically verified to be optimal---optimised over all possible measurements---for GHZ states \cite{Chaves2013}. Nevertheless, although this suggests that either parity measurement is suboptimal or the above interrogation time $t$ dependence should be improved in the parity-based scenario, \eqnref{eq:parity_bound} suffices to prove the superclassical precision scaling, $1/N^{5/3}$, as well as the $(\gamma\omega^2)^{1/3}$ behaviour of the asymptotic coefficient.

\section{Note on vanishing parameter value}
\label{app.omega_zero}

For $\omega\!=\!0$, both the GHZ-achievable bound \eref{eq.prevbound} and the OATSS-based expression \eref{eq.squeezenonzeroperpscaling} vanish. This does not mean that the precision is unbounded for the two cases, but rather suggests that the results give no information in such a limit. It is therefore not clear, what precision scaling can then be achieved.

In general, for the channel described by \eqnref{map} at $\omega \!=\! 0$, we get $\xi\!=\!1$, $\chi\!=\!0$, and $\partial\xi/\partial\omega \!=\! 0$, $\partial\chi/\partial\omega \!=\! (\ee^{-t \gamma }-1)/\gamma$.
For a GHZ state \eref{eq.GHZ} and parity measurement \eqnref{eq.parity}, one can show utilising \eqnref{eq.freqerrprop} that
\begin{equation}
\label{eq.GHZzeroscaling}
\Delta^2 \omega T  \underset{N\to\infty}{=} \frac{t \gamma ^2}{(1-\ee^{-t\gamma})^2}\, \frac{1}{N^2} ,
\end{equation}
for fixed $t$. This is minimised at $t_\t{opt} \!=\! \kappa/\gamma$, where $\kappa$ is a numerical constant. Similarly, at $\omega\!=\! 0$ for an OATSS along $y$ squeezed in $x$ [as in scenario (b) of \figref{fig.magnetosetup} of the main text] with squeezing parameter $\mu = (N/4)^{-2/3}$ one finds
\begin{equation}
\Delta^2 \omega T  \underset{N\to\infty}{=} \frac{5}{3\times 2^{2/3}} \frac{t \gamma ^2 }{(1-\ee^{-t\gamma})^2}\, \frac{1}{N^{5/3}} ,
\end{equation}
and again the optimal time is $t_\t{opt} = \kappa/\gamma$.

Thus, on the one hand, the local estimation approach which we employ above indicates that an improved scaling, even reaching the Heisenberg limit for GHZ states, is possible at the special $\omega\!=\!0$ parameter value.
Even if the value of $\omega$ is \emph{a priori} nonzero, one might then think that the precision scaling can be improved by adopting an iterative, adaptive strategy \cite{Wiseman2009,Berry2000,Hentschel2010}. By applying a bias (e.g., in the case of magnetometry, a magnetic field in opposite direction to the estimated field) to decrease the parameter after obtaining its first estimate, a better estimate is obtained with a precision which less heavily constrained by bounds Eqs.~\eref{eq.prevbound} and \eref{eq.squeezenonzeroperpscaling}, due to the lower effective value of $\omega$. On the other hand, the prior information on $\omega$ required to adjust the bias may scale prohibitively. We can compute the estimated mean-squared error for GHZ states and parity measurements (see above) and expand it in $omega$ to obtain $\Delta^2\omega T \approx c(\gamma,t) / N^2 + O(\omega^2)$, with $c$ given by \eqnref{eq.GHZzeroscaling}. For the Heisenberg-scaling term to dominate for a fixed $t$, the higher-order terms in the expansion must be negligible in comparison. However, we find (for even terms, odd terms vanish) that the $k$th term scales as $\omega^k N^{k-3}$, which implies that we need $\omega \ll N^{-(k-1)/k}$ to neglect the higher-order terms. For this to hold for all $k$, $\omega \ll 1/N$, which means that the prior information on $\omega$ must already be Heisenberg limited, as in the case of a decoherence-free local estimation scenario \cite{Jarzyna2014,Kolodynski2014}.
At first sight, this may indicate that such an adaptive scheme may not be successful for any prior distribution of finite width, and that
the value of $\omega$ must be perfectly known and set to zero for the above improved scalings to be observed.
However, recent results \cite{Macieszczak2014}, based on the Bayesian approach to estimation, indicate that
in the decoherence-free case the Heisenberg scaling is attained irrespectively of the prior knowledge of $\omega$.
Hence, we expect the transversal-noise model to behave similarly due to its decoherence-free-like regime at short interrogation times,
which would then prove the above adaptive strategy to also be efficient.

\section{Bound for scenario (a) of \figref{fig.magnetosetup}}
\label{app.scenarioabound}

We give below an analytical proof of \eqnref{eq.squeezenonzeroperpscalingA} in the limit $\omega\!\to\!0$ and strong numerical evidence for arbitrary $\omega$.
Following the method outlined at the beginning of \secref{sub.varscaling}, we find expressions for $\Delta^2 \omega_\t{(a)} T$ and $\Delta^2 \omega_\t{(b)} T$. We are looking for a lower bound on the former. From the expressions it can be seen that, in the case $\omega\!=\! 0$
\begin{equation}
(\Delta^2 \omega_\t{(a)} T) - e^{-2t\gamma} (\Delta^2 \omega_\t{(b)} T) = \frac{\gamma ^2 t \,\coth\!\left(\frac{\gamma  t}{2}\right) \cos\!\left(\frac{\mu }{2}\right)^{2-2N}}{N} .
\end{equation}
Now, since both the prefactor $e^{-2t\gamma}$ and $\Delta^2 \omega_\t{(b)} T$ are positive, any lower bound on this quantity is also a lower bound on $\Delta^2 \omega_\t{(a)} T$. The $t$-dependent factor $t \coth\!\left(\frac{\gamma  t}{2}\right)$ is lower bounded by $2/\gamma$ (attained when $t\!\to\!0$), while $\cos\!\left(\frac{\mu }{2}\right)^{2-2N}$ is lower bounded by 1. It follows that $(\Delta^2 \omega_\t{(a)} T)$ is lower bounded by $2\gamma/N$ as desired.

\begin{figure}[b!]
\centering
\includegraphics[width=0.95\columnwidth]{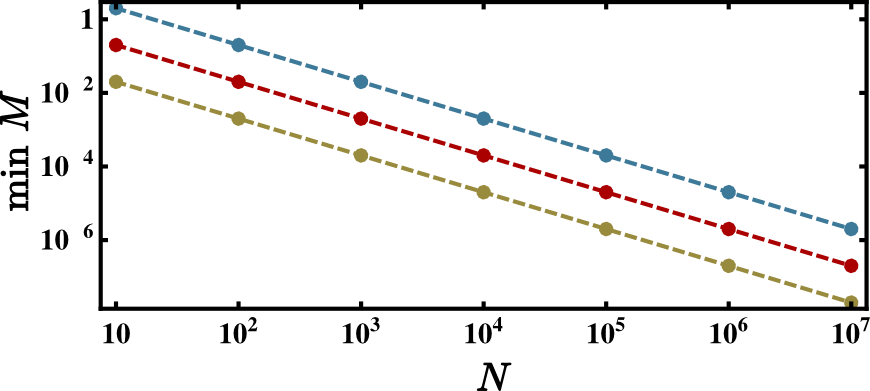}
\caption{(\emph{Circles}):~Results of minimising $M$ in \eqnref{eq:M} over $\mu$, $t$ versus $N$ for $\{\gamma,\omega\} \!=\! \{10,0.03\}$, $\{1.0,0.3\}$ and $\{0.1,0.03\}$ (top to bottom). (\emph{Dashed lines}):~Plots of $2\gamma/N$ for the same values of $\gamma$.}
\label{fig.scenarioabound}
\end{figure}

When $\omega\!>\!0$, the expressions for $\Delta^2 \omega_\t{(a)} T$ and $\Delta^2 \omega_\t{(b)} T$ become significantly more complicated. However, we can again look at a quantity
\begin{equation}
M = (\Delta^2 \omega_\t{(a)} T) - \left(\frac{A-B}{A+B}\right)^2 \, (\Delta^2 \omega_\t{(b)} T) ,
\label{eq:M}
\end{equation}
where
\begin{align}
A & = \cosh\!\left( \frac{1}{2} t \sqrt{\gamma^2 - 4\omega^2} \right), \\
B & = \gamma \sinh\!\left( \frac{1}{2} t \sqrt{\gamma^2 - 4\omega^2} \right) / \sqrt{\gamma^2 - 4\omega^2} .
\end{align}
Since $(A-B)^2/(A+B)^2$ and $\Delta^2 \omega_\t{(b)} T$ are positive, a lower bound on $M$ is again also a lower bound on $\Delta^2 \omega_\t{(a)} T$. We have not been able to prove an analytical bound for $M$, but for given values of $\omega$, $\gamma$, $N$ we can numerically minimise $M$ over $\mu$ and $t$. In \figref{fig.scenarioabound} we plot the results of several such minimisations. As can be seen, the numerics give very clear evidence that $\min M \!=\! 2\gamma/N$ and hence $\Delta^2 \omega_\t{(a)} T$ is lower bounded by $2\gamma/N$ as claimed.



%

\end{document}